\documentclass[aps,prb,superscriptaddress,twocolumn,showpacs]{revtex4}

\usepackage{amsmath, amsthm, amssymb}
\usepackage{graphicx}
\usepackage{color}

\usepackage{natbib}
\usepackage{hyperref}
\hypersetup{
        colorlinks=true,
}

\usepackage{tikz}
\usetikzlibrary{calc}
\usetikzlibrary{backgrounds}
\usetikzlibrary{arrows}
\usetikzlibrary{shapes.arrows}
\usetikzlibrary{decorations.markings}

\usepackage{verbatim}

\makeatletter
\@ifundefined{textcolor}{}
{%
 \definecolor{BLACK}{gray}{0}
 \definecolor{WHITE}{gray}{1}
 \definecolor{RED}{rgb}{1,0,0}
 \definecolor{GREEN}{rgb}{0,1,0}
 \definecolor{BLUE}{rgb}{0,0,1}
 \definecolor{CYAN}{cmyk}{1,0,0,0}
 \definecolor{MAGENTA}{cmyk}{0,1,0,0}
 \definecolor{YELLOW}{cmyk}{0,0,1,0}
}

\makeatother
\begin{document}

\title{Interplay of disorder and interaction in Majorana quantum wires}

\author{Alejandro M. Lobos}

\affiliation{Condensed Matter Theory Center and Joint Quantum Institute, Department
of Physics, University of Maryland, College Park, Maryland 20742-4111,
USA.}

\author{Roman M. Lutchyn}

\affiliation{Station Q, Microsoft Research, Santa Barbara, CA 93106-6105.}

\author{S. Das Sarma}

\affiliation{Condensed Matter Theory Center and Joint Quantum Institute, Department
of Physics, University of Maryland, College Park, Maryland 20742-4111,
USA.}

\date{\today}

\begin{abstract}
We study the interplay between disorder and interaction in one-dimensional
topological superconductors which carry localized Majorana zero-energy
states. Using Abelian bosonization and the perturbative renormalization
group (RG) approach, we obtain the RG-flow and the associated scaling
dimensions of the parameters and identify the critical points of the
low-energy theory. We predict a quantum phase transition from a topological
superconducting phase to a non-topological localized phase, and obtain
the phase boundary between these two phases as a function of the electron-electron
interaction and the disorder strength in the nanowire. Based on an instanton analysis which incorporates the effect of disorder, we also identify
a large regime of stability of the Majorana-carrying topological phase in the parameter
space of the model.
\end{abstract}
\maketitle
\textit{Introduction.} The search for topological phases of matter
has become an active and exciting pursuit in condensed matter physics~\cite{Wilczek'09}.
Among the many important examples of such phases are topological superconductors (SC)
supporting zero-energy Majorana bound states (MBS)~\cite{read_prb'00, kitaev2001, DasSarma_PRB'06,Fu08,zhang_prl'08,Sato_PRB09,Sau2010,Alicea_PRB10,Lutchyn2010,Oreg2010}.
A particularly promising realization of topological superconductivity
is one-dimensional (1D) semiconductor/SC heterostructures~\cite{Lutchyn2010, Oreg2010}.
In addition to being one of the simplest examples of fractionalization,
zero-energy MBS quasiparticles have Ising-like non-Abelian braiding
properties~\cite{Ivanov_PRL'01, Read09, Aliceaetal'10, Bonderson11} and can be used for topological
quantum computation~\cite{Nayak08_RMP_Topological_quantum_computation}.

The distinct feature of topological SCs is the ground-state
degeneracy due to the fermion parity encoded in the exponentially
localized zero-energy MBS~\cite{kitaev2001, Cheng09_Splitting_of_MF_in_p-wave_SC}. In a finite-length 1D wire,
this degeneracy is approximate and there is an exponentially small
energy splitting $e^{-L/\xi}$ due to a finite overlap of MBS.
Here $L$ and $\xi$ are the length of the wire and superconducting coherence
length, respectively. The presence of impurities in 1D p-wave SCs with broken time-reversal and spin $SU(2)$ symmetry (class D) ~\cite{Altland1997} adversely
affects the stability of the topological phase and drives a transition to a non-topological insulator phase~\cite{Motrunich01_Disorder_in_topological_1D_SC,  Brouwer00_Localization_Dirty_SC_wire,Gruzberg05_Localization_in_disordered_SC_wires_with_broken_SU2_symmetry, Brouwer11_Topological_SC_in_disorder_wires, Brouwer11_Probability_distribution_of_MFS_in_disordered_wires, Potter2010, Lutchyn2011, Stanescu11_MFs_in_SM_nanowires,Sau11_MF_in_real_materials}. The aforementioned QPT transition between topological and non-topological (localized) thermal insulator phases is accompanied by the change of the ground-state degeneracy splitting from exponential to algebraic in $L$~\cite{Brouwer11_Topological_SC_in_disorder_wires}. In other words, increasing the disorder strength leads to a topological quantum phase transition (QPT) from the
Majorana-carrying topological SC phase with quantum degeneracy
to a trivial phase with no end-MBS in the wire. The effect of electron-electron interactions in the
disordered SC wires have {\it not} been taken account before. The latter may have important implications for the topological
phase, and there may be QPTs associated with the tuning of the interaction
strength. Indeed, it is well known that the low-energy properties of 1D conductors are
strongly affected by both electron-electron interactions and disorder
\cite{Giamarchi_book}. Clarification of their combined effect
is crucial for our complete understanding of the topological phase
diagram of the system and ultimately for the experimental realization
of Majorana quantum wires in the laboratory~\cite{Mourik12_Signatures_of_MF}, where both
disorder and interactions would be inevitably present.

In this Letter, we investigate an important question concerning the
effect of both disorder and interaction on the stability of the topological
phase and go beyond the non-interacting results of Refs.\cite{Motrunich01_Disorder_in_topological_1D_SC,  Brouwer00_Localization_Dirty_SC_wire,Gruzberg05_Localization_in_disordered_SC_wires_with_broken_SU2_symmetry, Brouwer11_Topological_SC_in_disorder_wires, Brouwer11_Probability_distribution_of_MFS_in_disordered_wires, Potter2010, Lutchyn2011, Stanescu11_MFs_in_SM_nanowires,Sau11_MF_in_real_materials}, and of Refs. ~\cite{Gangadharaiah11_Majorana_fermions_in_1D_interacting_wires,Sela11_MF_in_strongly_interacting_helical_liquids,LutchynFisher'11,Stoudenmire11_Interaction_effects_in_1D_wires_with_Majorana_fermions},
where the effects of interaction have been studied in clean nanowires.
We consider a generic 1D p-wave SC and include the effects
of both quenched disorder and interaction using Abelian bosonization
and the replica method~\cite{edwards_replica}. We derive a set of coupled renormalization-group
(RG) equations for the parameters of the model, obtaining in the process
the quantum phase diagram of the system. Using these results in combination with
an instanton analysis allows us to analyze the topological stability of MBS under the influence of both interaction and disorder.
In general, disorder
and repulsive interactions reinforce their detrimental effects on
the topological SC phase and tend to eliminate the exponentially-split
ground state MBS degeneracy associated with different fermion parity~\cite{Fidkowski11_Majorana_fermions_in_wires_without_LRO}.
However, for a sufficiently strong initial induced pairing $\Delta$,
we predict a stable topological phase at low temperatures, even in
the presence of disorder and interaction. Our results are relevant
to recent experiments on semiconductor nanowires with strong spin-orbit
and Zeeman interactions, proximity-coupled to a s-wave bulk SC
\cite{Mourik12_Signatures_of_MF}, whose low-energy Hamiltonian
was shown to reduce to an effective 1D spinless p-wave
SC \cite{Lutchyn2010, Oreg2010},
and shed light on the question of the stability of MBS in realistic
situations.

\textit{Theoretical model.} We start with a model for p-wave spinless
fermions in a clean, single channel conductor of length $L$ with
open boundary conditions. In that case, the Hamiltonian for the 1D
SC wire in the continuum is

\begin{align*}
H_{0}^{\left(1\right)}= & \int_{0}^{L}dx\;\psi^{\dagger}\left(-\frac{\partial_{x}^{2}}{2m}-\mu\right)\psi-\Delta\psi\left(\frac{i\partial_{x}}{k_{F}}\right)\psi+\text{H.c.},
\end{align*}
where $\hbar=1$, $\psi\left(x\right)$ is the fermionic annihilation
field operator, $m$ is the effective mass, $\mu$ is the chemical
potential and $\Delta$ is the p-wave pairing interaction. In absence
of interactions and disorder, the Hamiltonian $H_{0}^{\left(1\right)}$
can be straightforwardly diagonalized by the means of a standard Bogoliubov
transformation. %
However, introducing interactions considerably complicates the theoretical
description and a different approach is needed. We therefore start
from the limit $\Delta=0$, and linearize the spectrum $\xi_{k}=k^{2}/2m-\mu$
around the Fermi points $\pm k_{F}$. This allows to express the fermion
field $\psi\left(x\right)$ as a sum of right- and left-movers $\psi\left(x\right)=e^{ik_{F}x}\psi_{R}\left(x\right)+e^{-ik_{F}x}\psi_{L}\left(x\right)$,
and to introduce the standard Abelian bosonization procedure of Fermi
fields
$\psi_{r} =\frac{1}{\sqrt{2\pi a}}U_{r}e^{-i\left(r\phi-\theta\right)}$, where $\quad r=\{R\left(+\right),L\left(-\right)\}$, and
$a\sim k_{F}^{-1}$ is the short-distance cutoff of the continuum
theory,
The bosonic fields $\phi\left(x\right),\theta\left(x\right)$
are conjugate canonical variables obeying the commutation relation
$\left[\phi\left(x\right),\theta\left(y\right)\right]=i\pi\text{sign}\left(y-x\right)/2$,
and $U_{r}$ are the standard Klein factors \cite{Giamarchi_book}.
Physically, $\phi\left(x\right)$ represents slowly-varying
fluctuations in the electronic density $\rho\left(x\right)=\rho_{0}-\partial_{x}\phi\left(x\right)/\pi$,
and $\theta\left(x\right)$ is related to the SC order parameter through
the relation $-i\psi\left(x\right)\partial_{x}\psi\left(x\right)\propto\psi_{R}\left(x\right)\psi_{L}\left(x\right)\propto e^{i2\theta\left(x\right)}$,
where we have neglected less relevant higher-order terms in $\partial_{x}\theta\left(x\right)$.
A short-range interaction $H_{0}^{\left(2\right)}=g\int dx\;\psi_{R}^{\dagger}\left(x\right)\psi_{R}\left(x\right)\psi_{L}^{\dagger}\left(x\right)\psi_{L}\left(x\right)$
acquires a simple form in terms of the bosonic fields, and the Hamiltonian
$H_{0}=H_{0}^{\left(1\right)}+H_{0}^{\left(2\right)}$ is therefore
given by
\begin{align}
\! H_{0}\!&\!=\!\int\! dx\left[\frac{vK}{2\pi}\left(\partial_{x}\theta\right)^{2}\!+\!\frac{v}{2\pi K}\left(\partial_{x}\phi\right)^{2}\!+\!\frac{2\Delta}{\pi a}\sin\left(2\theta\right)\!\right]\!.\label{eq:H_0_bosonized}
\end{align}
For $\Delta=0$, Eq.  (\ref{eq:H_0_bosonized}) reduces to the Luttinger
liquid (LL) model \cite{Giamarchi_book}, which describes gapless
plasmon excitations in the wire propagating with velocity $v\simeq v_{F}$,
and is parametrized by the the dimensionless Luttinger parameter $K=\sqrt{\frac{1-g/\pi v_{F}}{1+g/\pi v_{F}}}$
representing repulsive (attractive) interactions for $K<1$($K>1$).
The hypothesis of a short-ranged interaction in $H_{0}^{\left(2\right)}$ requires
the presence of strong screening in the nanowire. In a realistic situation, such as the case of Ref.
[\onlinecite{Mourik12_Signatures_of_MF}],
we assume that this screening is provided both by electrons in the
semiconductor and by surrounding SC.
In Eq.~(\ref{eq:H_0_bosonized}) we have neglected the
umklapp scattering which would introduce an additional term $\sim\cos\left(2\phi-4k_{F}x\right)$,
since we assume a filling incommensurate with the lattice \cite{Gangadharaiah11_Majorana_fermions_in_1D_interacting_wires}.

As follows from the analysis of Eq. (\ref{eq:H_0_bosonized}) made below, the SC pairing $\Delta$ around $K\approx 1$ is relevant [see Eq. (\ref{eq:RG_y_Delta})], and flows to strong coupling. Thus, at large enough $\Delta$, the field $\theta\left(x\right)$ is pinned to
the minima of $\sin2\theta$ and the SC state
breaks $\mathbb{U}(1)$ symmetry down to $\mathbb{Z}_{2}$. In the infinite system $L\rightarrow \infty$, the latter corresponds to two degenerate minima at $\theta\left(x\right)=-\pi/4,\;3\pi/4$ which are related to each other by
the global $\mathbb{Z}_{2}$ transformation $\theta\rightarrow\theta+\pi$~\cite{Fidkowski11_Majorana_fermions_in_wires_without_LRO}. Such a transformation is implemented by the fermion parity
operator $P=\left(-1\right)^{N_{F}}=\exp\left[-i\int_{0}^{L}\partial_{x}\phi\left(x\right)dx\right]$ with $N_{F}$ the total fermion number operator. The degenerate ground states characterized by different fermion parity read $\left|\text{even/odd}\right\rangle =\left(\left|-\pi/4\right\rangle \pm\left|3\pi/4\right\rangle \right)/\sqrt{2}$. In the case of a large but finite $L$,
the two degenerate groundstates are
split in energy due to quantum tunneling between the two minima $\theta\left(x\right)=-\pi/4,\;3\pi/4$. The splitting energy can be calculated using instanton analysis $\delta E = A_{f}e^{-S_{\text{inst}}}$, where $S_{\text{inst}}$
is the action of the Euclidean instanton $\theta_{0}\left(x,\tau\right)$
(where $\tau$ is the imaginary-time), obeying the boundary conditions
$\theta_{0}\left(x,-\infty\right)=-\pi/4$ and $\theta_{0}\left(x,\infty\right)=3\pi/4$,
and $A_{f}$ is a prefactor due to quantum fluctuations around those minima
\cite{Coleman_instantons}. The instanton configuration minimizing $S_{\rm inst}$ is spatially
uniform rendering effectively a $0+1$ dimensional
problem, whose corresponding action is \cite{Fidkowski11_Majorana_fermions_in_wires_without_LRO}
\begin{align}
S_{\text{inst}} & =\frac{4\sqrt{K}}{\pi}\frac{L}{\xi},\label{eq:S_inst}
\end{align}
with $\xi=v/\Delta$ the SC coherence length. The instanton-analysis
therefore predicts an energy splitting scaling as $\delta E\propto \exp(-\frac{4\sqrt{K}}{\pi}\frac{L}{\xi})$, in agreement with the non-interacting Majorana chain~\cite{kitaev2001}.

We now introduce quenched disorder into model (\ref{eq:H_0_bosonized}).
We consider the case of a short-range Gaussian disorder potential
$V\left(x\right)$ that couples to the fermionic density, $H_{\text{dis}}=-\int dx\; V\left(x\right)\rho\left(x\right)$
and characterized by $\left\langle V\left(x\right)V\left(y\right)\right\rangle =D_{b}\delta\left(x-y\right).$
In bosonized language, the disordered Hamiltonian is
\begin{eqnarray} 
H_{\text{dis}} & =&\int dx\left[-\eta\left(x\right)\frac{\partial_{x}\phi\left(x\right)}{2\pi}+\xi\left(x\right)\frac{e^{-i2\phi}}{2\pi a}+\text{H.c.}\right],\label{eq:H_dis_rho_projected}
\end{eqnarray}
where we have defined the disordered potentials $\eta\left(x\right)\equiv\frac{1}{N}\sum_{q\sim0}e^{iqx}V\left(q\right)$ and
$\xi\left(x\right)\equiv\frac{1}{N}\sum_{q\sim0}e^{iqx}V\left(q-2k_{F}\right)$.
The forward scattering term $-\eta\left(x\right)\partial_{x}\phi\left(x\right)/2\pi$
can be eliminated by the means of a gauge transformation $\phi\left(x\right)\rightarrow\phi\left(x\right)-\frac{K}{v}\int^{x}dy\;\eta\left(y\right)$,
reflecting the fact that forward scattering does not affect the thermodynamic
properties of the system \cite{Giamarchi1988}.
We next implement the replica method, that consists in introducing
the set of ``replicas'' of the system $\phi\left(x\right),\theta\left(x\right)\rightarrow \{\phi_{i}\left(x\right),\theta_{i}\left(x\right) \}$,
with $i=1,2,\dots,n$, allowing a simpler integration over different
disorder configurations \cite{Giamarchi_book, edwards_replica}.
After integrating out the Gaussian field  $V\left(x\right)$
the replicated action of the 1D system becomes
\begin{eqnarray}
S & =&\sum_{j=1}^{n} \int d\tau  \left[ \int dx\ \frac{\partial_{x}\phi_{j}}{i\pi}\dot{\theta}_{j}+ H_{0,j}\left(\tau\right) \right]-\sum_{i,j=1}^{n} \frac{D_{b}}{\left(2\pi a\right)^{2}} \nonumber \\
&&\times\int dx d\tau d\tau^\prime  \cos2\left[\phi_{i}\left(x,\tau\right)-\phi_{j}\left(x,\tau^{\prime}\right)\right],  \label{eq:S_replicated} 
\end{eqnarray}
where the Hamiltonian $H_{0,j}$ is defined in Eq.~\eqref{eq:H_0_bosonized}.
In the absence of SC pairing, this model was studied by Giamarchi and Schulz
in the context of the localization transition, predicted to occur
at the critical value $K_{c}=3/2$ for spinless fermions, in the limit of weak disorder
 \cite{Giamarchi1988}. For $K<K_{c}$, disorder
flows to strong coupling and the groundstate corresponds to a pinned
charge-density-wave (PCDW), characterized by a localization length
$\xi_{\text{loc}}\propto D_{b}^{1/\left(3-2K\right)}$. Above $K_{c}$,
the LL phase remains stable, describing a ``delocalized'' electronic
fluid. In the presence of SC pairing, the LL fixed-point is never stable, as we show below.

\textit{RG analysis.} The critical properties of model (\ref{eq:S_replicated})
can be studied in the framework of perturbative RG around the LL fixed-point. Following standard derivations \cite{Giamarchi_book, suzumura_mean_field},
we expand the partition function corresponding to action $S$ at first-order
in the small parameter $D_{b}$, and up to second order
in $\Delta$. We implement the RG procedure in real-space, which leaves invariant the
LL fixed-point Hamiltonian, and obtain the following system of RG-flow
equations
\begin{align}
\frac{dK\left(\ell\right)}{d\ell} & =y_{\Delta}^{2}\left(\ell\right)-K^{2}\left(\ell\right)y_{b}\left(\ell\right),\label{eq:RG_K}\\
\frac{dv\left(\ell\right)}{d\ell} & =-v\left(\ell\right)K\left(\ell\right)y_{b}\left(\ell\right),\label{eq:RG_v}\\
\frac{dy_{\Delta}\left(\ell\right)}{d\ell} & =\left[2-K^{-1}\left(\ell\right)\right]y_{\Delta}\left(\ell\right),\label{eq:RG_y_Delta}\\
\frac{dy_{b}\left(\ell\right)}{d\ell} & =\left[3-2K\left(\ell\right)\right]y_{b}\left(\ell\right),\label{eq:RG_y_delta_b}
\end{align}
where we have introduced the dimensionless variables $y_{\Delta}=2\Delta a/v$ and
$y_{b}=D_{b}a/4 \pi v^{2}$. Physically Eq.~\eqref{eq:RG_K} describes the renormalization
of interactions in the wire {[}parametrized by $K\left(\ell\right)${]}
induced by superconductivity and disorder. While $y_{\Delta}\left(\ell\right)$
couples to field $\theta\left(x\right)$,
favoring a SC ground state with broken $\mathbb{Z}_{2}$-symmetry,
the parameter $y_{b}\left(\ell\right)$ couples to the dual field
$\phi\left(x\right)$ and tries to pin the density to the disorder
potential, thus opposing a SC ground state. These competing effects
are reflected in the different signs of the prefactors in Eq.~(\ref{eq:RG_K}):
$y_{\Delta}\left(\ell\right)$ renormalizes
$K\left(\ell\right)$ to larger values, inducing attractive interactions
in the wire, while $y_{b}\left(\ell\right)$ drives $K\left(\ell\right)\rightarrow0$
enhancing the effect of repulsive interactions. In the limit $\left\{ y_{\Delta}\left(\ell\right),y_{b}\left(\ell\right)\right\} \rightarrow0$
the properties of the system are determined by the value of $K\left(\ell\right)$,
i.e., the coupling $y_{\Delta}\left(\ell\right)$
becomes relevant (in the RG sense) for $K\left(\ell\right)>1/2$, whereas $y_{b}\left(\ell\right)$ is relevant for $K\left(\ell\right)<3/2$~\cite{Giamarchi1988,suzumura_mean_field}.
From this RG-analysis we extract two important conclusions: 1) the
non-interacting limit $K=1$ is an unstable point in parameter-space,
and 2) repulsive interaction and disorder reinforce each other's detrimental
effects on the topological SC. 
\begin{figure}
\begin{centering}
\includegraphics[bb=130bp 0bp 550bp 200bp,clip,scale=0.6]{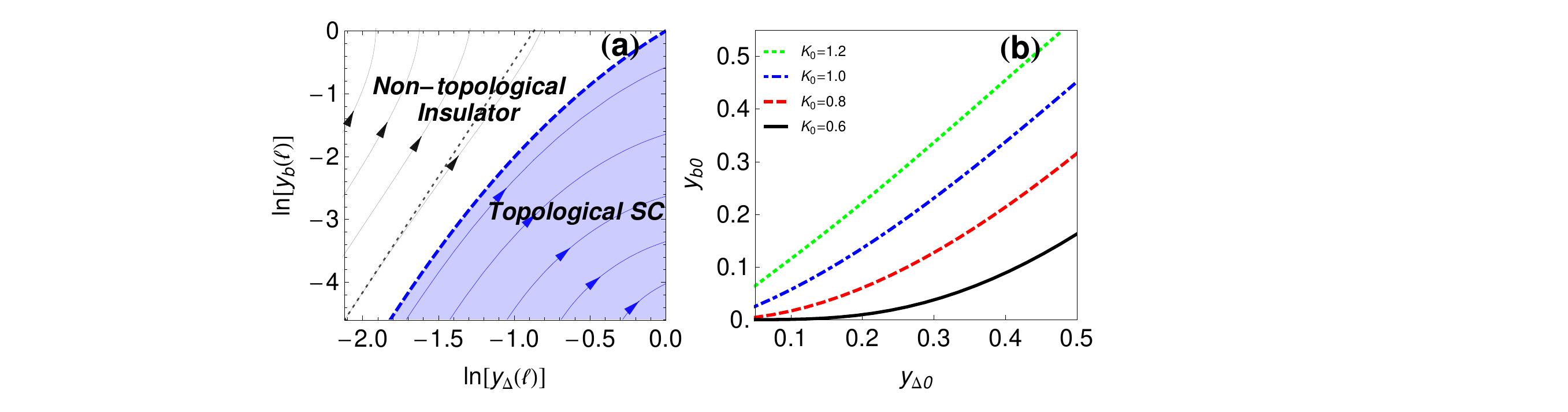}
\par\end{centering}

\caption{\label{fig:figure}(a) Parametric dependence of $y_{b}\left(\ell\right)$
vs $y_{\Delta}\left(\ell\right)$, as obtained from the numerical
solution of the RG-flow Eqs. (\ref{eq:RG_K})-(\ref{eq:RG_y_delta_b}),
for fixed initial parameters $K_{0}=0.65$ and $y_{s0}=0$ (log-log
scale). The thick dashed curve is the critical line, separating the
topological SC phase (shaded area) from the non-topological disordered
phase, and the thin dotted line is our analytical estimate $y_{b}\sim y_{\Delta}^{\nu}$,
valid in the limit $\left\{ y_{b}\left(\ell\right),y_{\Delta}\left(\ell\right)\right\} \rightarrow0$.
(b) Phase diagram in $y_{\Delta0},y_{b0}$ space obtained for $y_{s0}=0$
and different values of $K_{0}$. The curves correspond to the critical
lines $y_{b0}$ vs $y_{\Delta0}$, satisfying the condition $y_{\Delta}\left(\ell^*\right)=y_{b}\left(\ell^*\right)=1$.
The area below each curve represents the regime for which topological
SC is expected to dominate over disorder.}
\end{figure}
Note that within the experimentally
interesting regime $1/2<K\left(\ell\right)<3/2$ both $y_{\Delta}\left(\ell\right)$
and $y_{b}\left(\ell\right)$ are competing perturbations flowing
simultaneously to strong coupling. Moreover, in the non-interacting
case $K=1$, $y_{\Delta}\left(\ell\right)$ and $y_{b}\left(\ell\right)$
have the same scaling dimension. In order to maintain the internal
consistency of our perturbative approach, the RG flow has to be stopped
at a value $\ell^{*}$ for which one of the couplings reaches the
strong-coupling regime, i.e., $\max\left[y_{\Delta}\left(\ell^{*}\right),y_{b}\left(\ell^{*}\right)\right]=1$.
Although strictly speaking our approach is not applicable in the strong-coupling
regime, the fact that $\theta\left(x\right)$ and $\phi\left(x\right)$
are dual fields that cannot order simultaneously allows us to reasonably
conjecture that there are no intermediate fixed-points in the RG flow,
and therefore to classify the nature of the ground state according
to the coupling that first reaches the above condition \cite{suzumura_mean_field}. When the two
competing couplings reach the strong coupling regime simultaneously
[i.e., $y_{b}\left(\ell^{*}\right)=y_{\Delta}\left(\ell^{*}\right)=1$],
the system does not order and this condition defines a critical line
of QPTs that separates the topological SC phase with broken $\mathbb{Z}_{2}$
symmetry from the PCDW insulating phase (cf. thick dashed line in
Fig. \ref{fig:figure}(a)).

From the lowest-order RG equations one obtains the approximate solutions $y_{b}\left(\ell\right)=y_{b0}e^{\left(3-2K\right)\ell}$,
$y_{\Delta}\left(\ell\right)=y_{\Delta0}e^{\left(2-K^{-1}\right)\ell}$,
which together produce the relative scaling $y_{b}\sim y_{\Delta}^{\nu}$
with $\nu=\left(3-2K\right)/\left(2-K^{-1}\right)$. Physically, this
means that interactions (encoded in $\nu$) determine the scaling
of disorder strength relative to the SC order parameter: for $K>1$
(attractive interactions) disorder grows slower than SC, while the
inverse occurs for $K<1$ (repulsive interactions). In Fig. \ref{fig:figure}(a)
we show the parametric dependence of $y_{b}\left(\ell\right)$ as
a function of $y_{\Delta}\left(\ell\right)$, for the initial condition
$K_{0}=0.65$. The continuous lines correspond to
the numerical solution of Eqs. (\ref{eq:RG_K})-(\ref{eq:RG_y_delta_b}),
and the dotted line is our analytical result $y_{b}\sim y_{\Delta}^{\nu}$,
valid in the limit $\left\{ y_{b}\left(\ell\right),y_{\Delta}\left(\ell\right)\right\} \rightarrow0$.
At the phase boundary (thick dashed line), this result implies the
approximate relation $y_{b0}\sim y_{\Delta0}^{\nu}$ for the initial values,
which together with the relation: $D_{b}=2\pi v_{F}/\tau_{e}$ (where
$\tau_{e}$ is elastic scattering time), produces $1/\tau_{e}E_{F}\sim(\Delta/E_{F})^{\nu}$.
Interestingly, for $K=1$ we find that the critical condition for
the topological-non-topological transition is $1/\tau_{e}\sim \Delta$,
which exactly coincides with the results obtained in the non-interacting
case \cite{Motrunich01_Disorder_in_topological_1D_SC,Brouwer00_Localization_Dirty_SC_wire,Gruzberg05_Localization_in_disordered_SC_wires_with_broken_SU2_symmetry,Brouwer11_Topological_SC_in_disorder_wires,Brouwer11_Probability_distribution_of_MFS_in_disordered_wires,Sau11_MF_in_real_materials}.
Note, however, that in the interacting case the equation for the phase boundary involves an additional energy scale $E_F$ and has a non-trivial dependence on the electron-electron interactions.

The above procedure leads to a qualitative ``phase-diagram'' in
terms of the initial parameters of the model. In Fig. \ref{fig:figure}(b)
we plot the critical curves in $y_{\Delta0}$-$y_{b0}$ space, for
different initial values of interaction $K_{0}$. The area below each
curve represents the regime for a stable topological SC supporting
MBS. Starting from the inital value $K_{0}=0.6$ (i.e., strongly interacting
wire), note that the topological region expands as the interaction
becomes increasingly attractive.%

\textit{Topological stability of MBS.} To study the effect of interaction and
disorder on the stability of MBS, we evaluate the energy-splitting
$\delta E$ in the regime where $\Delta$ flows first to strong coupling.
As mentioned before, in order for the topological SC phase to be stable,
$\delta E$ should scale exponentially with $L$.
Our approach therefore consists in integrating the RG-flow up to the
scale $\ell^{*}=\ln\left(y_{\Delta0}\right)/\left(K_0^{-1}-2\right)$
{[}i.e., such that $y_{\Delta}\left(\ell^{*}\right)=1${]}, and calculating there the instanton action $S_{\text{inst}}$ in
presence of the backscattering term in (\ref{eq:H_dis_rho_projected}).
Since in that regime $y_{b}\left(\ell^{*}\right)\ll1$, the effect
of backscattering can be accounted for perturbatively, and we can
make use of the instanton solution $\theta_{0}\left(\tau\right)$
found in the clean case. The contributions of backscattering to $S_{\text{inst}}$
can be divided into: a) an \textit{explicit} contribution, arising
from the presence of the term $\sim D_{b}\left(\ell^{*}\right)\left\langle \cos\left[2\phi\left(x,\tau_{1}\right)-2\phi\left(x,\tau_{2}\right)\right]\right\rangle $ in the action,
and b) an \textit{implicit} contribution, originated in the indirect effect of  $y_{b}\left(\ell\right)$ on the other couplings through the RG-flow equations.
Since in the regime of interest
$\Delta$ ``locks'' the phase $\theta$ to the minima of the $\sin2\theta$
potential, $\phi$ becomes a strongly fluctuating field and therefore
the contribution a) is strongly suppressed, i.e. it scales as $\left\langle \cos\left[2\phi\left(x,\tau\right)-2\phi\left(x,0\right)\right]\right\rangle \sim\exp\left(-\left|\tau E_F\right|L/\xi \right)$  [See Appendix \ref{appendix}].
This constitutes a subleading correction to $S_{\text{inst}}$ which is neglected in the following analysis.
We therefore focus on the more important contribution b). The expression of the instanton
action $S_{\text{inst}}\left(\ell^{*}\right)$ is formally identical
to Eq. (\ref{eq:S_inst}) with the change $K\rightarrow K\left(\ell^{*}\right)$.
Integrating RG-flow Eq. (\ref{eq:RG_K}) up to the scale $\ell^{*}$ yields
 (at lowest order in the parameters $y_{\Delta}$ and $y_{b}$)
$K\left(\ell^{*}\right)=K_{\text{cl}}-\delta K_{\text{dis}},$
where $K_{\text{cl}}=K_{0}+K_0\left(4 K_0-2 \right)^{-1}$
is the renormalized Luttinger parameter in the clean limit $l_{e}=v\tau _e \rightarrow\infty$,
and where $\delta K_{\text{dis}}=K_{0}^{2} \left(3-2K_{0}\right)^{-1} \left(k_Fl_{e}\right)^{-1} \left(k_F\xi/2\right)^{\nu}$
 is the effect of disorder [See Appendix \ref{appendix}]. Replacing $K\left(\ell^*\right)$ into (\ref{eq:S_inst}) yields
\begin{eqnarray}
S_{\text{inst}} & =&\frac{4\sqrt{K_{\text{cl}}}}{\pi}\left[\frac{L}{\xi}-\frac{L}{2 l_e} \frac{K_{0}^{2}}{K_{\text{cl}}\left(3-2K_{0}\right)}\left(\frac{k_F\xi}{2}\right)^{\nu-1}\right].\label{eq:S_inst_strong_coupling}
\end{eqnarray}
This result encodes the interplay of interaction and (weak) disorder on the topological degeneracy
of MBS through the relation $\delta E \propto e^{-S_{\text{inst}}\left( \ell^*\right)}$, and constitutes an important generalization of
the non-interacting results in Ref. \cite{Brouwer11_Probability_distribution_of_MFS_in_disordered_wires} to the interacting case.
Physically, it expresses the fact that MBS are stable as long as disorder is weak, such that $\xi \left(k_F\xi\right)^{\nu-1} \ll l_{e}$. Note that the internal consistency of the bosonization approach requires the energy cutoff $\Lambda_0=v_F k_F$ to be much larger than $\Delta$. This implies that $k_F \xi \gg 1$ and we therefore conclude that effect of disorder on MBS energy splitting is enhanced (lessened) for repulsive (attractive) interactions, which is one of the main results of this paper. Interestingly, one can notice that the non-interacting results of Ref. \cite{Brouwer11_Probability_distribution_of_MFS_in_disordered_wires} are recovered for $K_{0}=1$ and $\nu=1$.
While Eq. (\ref{eq:S_inst_strong_coupling}) is only valid in the regime
$1/2<K_{0}<3/2$ due to the lowest-order approximation in
the integration of the RG-flow, a numerical integration of Eqs. (\ref{eq:RG_K})-(\ref{eq:RG_y_delta_b})
allows to generalize it to any $K_{0}$.

\textit{Conclusions. }We have carried out a RG analysis of the topological
superconductivity in a 1D p-wave SC wire
in the presence of both electron-electron interaction and disorder, treating them on equal footing.
Our results provide useful insights into their interplay and
are relevant to understand more realistic situations (e.g., Ref. \cite{Mourik12_Signatures_of_MF}).
The solution of the RG-flow Eqs. (\ref{eq:RG_K})-(\ref{eq:RG_y_delta_b}) combined with the calculation of the instanton action in Eq. (\ref{eq:S_inst_strong_coupling})
demonstrate that a topological SC state that supports
stable non-Abelian MBS could be in principle realized on a large regime of parameter space.

\begin{acknowledgments}
The authors are grateful to Chetan Nayak, Meng Cheng, Michael Levin,
So Takei and Thierry Giamarchi for valuable discussions. We acknowledge
support from DARPA QuEST, JQI-NSF-PFC and Microsoft Q.
\end{acknowledgments}

\appendix
\section{CALCULATION OF THE INSTANTON ACTION $S_{\text{inst}}$ IN PRESENCE OF WEAK DISORDER }\label{appendix}

In this supplementary document we present the details of the instanton calculation on the ground-state degeneracy splitting in the presence of weak disorder.

Let us consider the action of the system $S=S_{0}+S_{\text{dis}}$
{[}c.f. Eq. (4) in the manuscript{]}, in the strong-coupling regime
$y_{\Delta}\rightarrow1$:
\begin{eqnarray}
S_{0}^{*} & = & \int dxd\tau\left[\frac{\partial_{x}\phi}{i\pi}\dot{\theta}+\frac{v\left(\ell^{*}\right)K\left(\ell^{*}\right)}{2\pi}\left(\partial_{x}\theta\right)^{2}\right.\label{eq:S_0}\\
 &  & \left.+\frac{v\left(\ell^{*}\right)}{2\pi K\left(\ell^{*}\right)}\left(\partial_{x}\phi\right)^{2}+\frac{2\Delta\left(\ell^{*}\right)}{\pi\xi}\sin2\theta\right]\nonumber \\
S_{\text{dis}}^{*} & = & -\frac{D_{b}\left(\ell^{*}\right)}{\left(2\pi\xi\right)^{2}}\int_{x\tau_{1}\tau_{2}}\cos\left[2\phi\left(x,\tau_{1}\right)-2\phi\left(x,\tau_{2}\right)\right],\label{eq:S_dis}
\end{eqnarray}
where $\ell^{*}$ is the scale at which the strong-coupling condition
$y_{\Delta}\left(\ell^{*}\right)=1$ is reached, and where we have
dropped the replica indices since we only keep the diagonal terms
[i.e., at the lowest order of perturbation theory in $D_b$, only diagonal terms contribute].
In this limit, $S_{0}^{*}\equiv S_{0}\left(\ell^{*}\right)$ is dominated
by the term $\sim\Delta\left(\ell^{*}\right)\sin2\theta\left(x,\tau\right)$
and the field $\theta\left(x,\tau\right)$ is pinned around the classical
values $\theta_{0}=\left\{ -\pi/4,3\pi/4\right\} $ \cite{Fidkowski11_Majorana_fermions_in_wires_without_LRO}.
These minima are connected to each other via the classical instantons
$\theta_{\text{inst}}\left(\tau\right)=\frac{\pi}{2}+2\arctan\left[\tanh\left(\tau/\tau_{0}\right)\right]$,
with $\tau_{0}=\sqrt{K\left(\ell^{*}\right)}\xi/v\left(\ell^{*}\right)$
the unit of time \cite{Coleman_instantons}. The goal is to calculate
the instanton action $S_{\text{inst}}$ in the presence of the disorder
term. Since in this limit $S_{\text{dis}}^{*}\equiv S_{\text{dis}}\left(\ell^{*}\right)$
is a perturbation to $S_{0}^{*}$, we can obtain the lowest-order
effects by injecting the instanton $\theta_{\text{inst}}\left(\tau\right)$,
obtained in absence of disorder, back into action $S^{*}=S_{0}^{*}+S_{\text{dis}}^{*}$.
The difficulty of this procedure resides in the fact that $S_{\text{dis}}^{*}$
is expressed in terms of the dual field $\phi\left(x,\tau\right)$.
To overcome this problem we first expand the function $\sin2\theta\left(x,\tau\right)$
around the minima $\theta_{0}$, and obtain the Gaussian approximation

\begin{eqnarray}
S_{0}^{*} & \simeq & \int dxd\tau\left[\frac{\partial_{x}\phi}{i\pi}\dot{\theta}+\frac{v\left(\ell^{*}\right)K\left(\ell^{*}\right)}{2\pi}\left(\partial_{x}\theta\right)^{2}\right.\nonumber \\
 &  & \left.+\frac{v\left(\ell^{*}\right)}{2\pi K\left(\ell^{*}\right)}\left(\partial_{x}\phi\right)^{2}+\frac{4\Delta\left(\ell^{*}\right)}{\pi\xi}\theta^{2}\right]\label{eq:S_0_Gaussian}
\end{eqnarray}
This approximation enables to perform analytical calculations around
the strong-coupling limit $y_{\Delta}\rightarrow1$.

We then focus on $S_{\text{dis}}^{*}$ and expand the backscattering
term $\sim\cos\left[2\phi\left(x,\tau_{1}\right)-2\phi\left(x,\tau_{2}\right)\right]$
in powers of the time-derivative $\dot{\phi}\left(x,\tau\right)$,
in order to extract the effects of disorder on the parameters $\Delta\left(\ell^{*}\right)$
and $K\left(\ell^{*}\right)$ of $S_{0}^{*}$. Note, however, that since
$\theta\left(x,\tau\right)$ is pinned, $\phi\left(x,\tau\right)$
is a strongly fluctuating field, and consequently one needs to introduce
the normal-order to perform the expansion safely \cite{giamarchi_book_1d},
i.e., $\cos\left[2\phi\left(x,\tau_{1}\right)-2\phi\left(x,\tau_{2}\right)\right]=:\cos\left[2\phi\left(x,\tau_{1}\right)-2\phi\left(x,\tau_{2}\right)\right]:\exp\left[-\frac{1}{2}\left\langle \left[2\phi\left(x,\tau_{1}\right)-2\phi\left(x,\tau_{2}\right)\right]^{2}\right\rangle _{0}\right],$
where ``$:\ :$'' denotes normal-ordering and the average
$\left\langle \dots\right\rangle _{0}$ is taken with respect to Eq.
(\ref{eq:S_0_Gaussian}). Introducing the center-of-mass and relative
coordinates $\tau=\left(\tau_{1}+\tau_{2}\right)/2$ and $\tau_{r}=\tau_{1}-\tau_{2}$,
we can express $S_{\text{dis}}^{*}$ as
\begin{align}
S_{\text{dis}}^{*} & \simeq-\frac{D_{b}\left(\ell^{*}\right)}{\left(2\pi\xi\right)^{2}}\int dxd\tau d\tau_{r}\;\left[1-2\left(\dot{\phi}\left(x,\tau\right)\right)^{2}\tau_{r}^{2}\right]\nonumber \\
 & \times e^{-\frac{1}{2}\left\langle \left[2\phi\left(x,\tau_{r}\right)-2\phi\left(x,0\right)\right]^{2}\right\rangle _{0}}\label{eq:S_dis_approx}
\end{align}
The first term in the square bracket yields a constant term and is
of no interest to us. The other term $\sim\left(\dot{\phi}\left(x,\tau\right)\right)^{2}$
couples to the field $\theta\left(x,\tau\right)$, and consequently,
to the instanton $\theta_{\text{inst}}\left(\tau\right)$. This can
be seen directly from the equation of motion for $\phi\left(x,\tau\right)$

\begin{eqnarray}
\dot{\phi}\left(x,\tau\right) & = & \left[H_{0}^{*}\left(\tau\right),\phi\left(x,\tau\right)\right]\nonumber \\
 & = & -iv\left(\ell^{*}\right)K\left(\ell^{*}\right)\nabla\theta\left(x,\tau\right)+\frac{i2\pi\Delta\left(\ell^{*}\right)}{\xi}\nonumber \\
 &  & \times\int dx^{\prime}\text{sgn}\left(x^{\prime}-x\right)\sin2\theta\left(x^{\prime},\tau\right).\label{eq:Eq_of_motion_phi}
\end{eqnarray}
We can now evaluate the corrected single-instanton action by replacing
the Eq. (\ref{eq:Eq_of_motion_phi}) into (\ref{eq:S_dis_approx}),
and computing $S_{\text{inst}}^{*}=S_{0}^{*}\left[\theta_{\text{inst}}\right]+S_{\text{dis}}^{*}\left[\theta_{\text{inst}}\right]$,
where we have injected the classical instanton solution $\theta_{\text{\text{inst}}}\left(\tau\right)$.
Here

\begin{eqnarray}
S_{0}^{*}\left[\theta_{\text{inst}}\right] & = & \frac{4\sqrt{K\left(\ell^{*}\right)}}{\pi}\frac{L}{\xi}\label{eq:S_0_str_coupling}\\
S_{\text{dis}}^{*}\left[\theta_{\text{inst}}\right] & = & -\frac{8D_{b}\left(\ell^{*}\right)\tau_{0}\left(\ell^{*}\right)}{3\xi^{2}}\left(\frac{\Delta\left(\ell^{*}\right)}{\xi}\right)^{2}\int_{-L/2}^{L/2}dx\ x^{2}\nonumber \\
 &  & \times\int d\tau_{r}\;\tau_{r}^{2}e^{-\frac{1}{2}\left\langle \left[2\phi\left(x,\tau_{r}\right)-2\phi\left(x,0\right)\right]^{2}\right\rangle _{0}},\label{eq:S_dis_str_coupling}
\end{eqnarray}
where the result $\int dz\;\sin^{2}\left(2\theta_{\text{inst}}\left(z\right)\right)=4/3$
has been used in Eq. (\ref{eq:S_dis_str_coupling}). The correlator
in the exponential is evaluated using stantard techniques \cite{giamarchi_book_1d}
and we obtain the expression
\begin{align*}
\left\langle \left[\phi\left(\tau\right)-\phi\left(0\right)\right]^{2}\right\rangle _{0} & =2\pi\left[\frac{x^{2}}{L}+\frac{L}{\pi^{2}}\right]\sqrt{\frac{\Delta\left(\ell^{*}\right)K\left(\ell^{*}\right)}{v\left(\ell^{*}\right)\xi}}f\left(\tau\right),
\end{align*}
with $f\left(\tau\right)\equiv1-\exp\left(-2\left|\tau\right|/\tau_{0}\right)$.
Replacing $\Delta\left(\ell^{*}\right)=v\left(\ell^{*}\right)/\xi$,
and $D_{b}\left(\ell^{*}\right)=v^{2}\left(\ell^{*}\right)/l_{e}$
in the above expression, we obtain
\begin{align}
S_{\text{dis}}^{*}\left[\theta_{\text{inst}}\right] & =-\frac{8K^{2}\left(\ell^{*}\right)}{3}\left(\frac{L^{3}}{l_{e}\xi^{2}}\right)A\left(\frac{L}{\xi}\right),\label{eq:S_dis_2}
\end{align}
where we have defined the function $A\left(\gamma\right)\equiv\int dz\int_{-1/2}^{1/2}dy\; z^{2}y^{2}\exp\left[-4\gamma\sqrt{K\left(\ell^{*}\right)}\left(\frac{1}{\pi}+\pi y^{2}\right)f\left(z\right)\right]$.
The analytical form of $A\left(\gamma\right)$ is not particularly
illuminating and we rather point out the scaling property $A\left(\gamma\right)\propto1/\gamma^{3}$,
valid at large $\gamma$. This yields
\begin{align}
S_{\text{dis}}^{*}\left[\theta_{\text{inst}}\right] & \propto-\frac{\xi}{\ell_{e}},\label{eq:S_dis_final}
\end{align}
which indicates that the perturbative term $S_{\text{dis}}^{*}\left[\theta_{\text{inst}}\right]$
does not scale with the size of the system, and therefore can be dropped
in the thermodynamical limit $L/\xi\rightarrow\infty$ in front of
$S_{\text{0}}^{*}\left[\theta_{\text{inst}}\right]$ in Eq. (\ref{eq:S_0_str_coupling}).

We now proceed to estimate the renormalized parameter $K\left(\ell^{*}\right)$.
The zeroth-order approximation $K\left(\ell\right)\simeq K_{0}$ allows
to integrate straightforwardly the RG-flow Eqs. (7) and (8) in the
manuscript, producing respectively $y_{\Delta}\left(\ell\right)=y_{\Delta0}e^{\left(2-K_{0}^{-1}\right)\ell}$
and $y_{b}\left(\ell\right)=y_{b0}e^{\left(3-2K_{0}\right)\ell}$,
and to estimate the maximal scale $\ell^{*}=\ln\left(y_{\Delta0}\right)/\left(K^{-1}-2\right)$
from the strong-coupling condition $y_{\Delta}\left(\ell^{*}\right)=1$.
Replacing these results back into Eq. (5) in the manuscript yields
the lowest-order correction of parameter $K\left(\ell^{*}\right)$
\begin{eqnarray*}
K\left(\ell^{*}\right) & = & K_{0}+\int_{0}^{\ell^{*}}d\ell^{\prime}\frac{dK\left(\ell^{\prime}\right)}{d\ell},\\
 & \simeq & K_{0}+\frac{1}{4-2K_{0}^{-1}}-\frac{K_{0}^{2}}{3-2K_{0}}\frac{y_{b0}}{\left(y_{\Delta0}\right)^{\nu}},
\end{eqnarray*}
where $\nu=\left(3-2K_{0}\right)/\left(2-K_{0}^{-1}\right)$ and where
the property $\left\{ y_{\Delta0},y_{b}\right\} \ll1$ has been used.
This result, along with the relations $\Delta_{0}=v_{0}/\xi$, and
$D_{b0}=v_{0}^{2}/l_{e}$, are used to define $K\left(\ell^{*}\right)=K_{\text{cl}}-\delta K_{\text{dis}}$,
where:

\begin{eqnarray}
K_{\text{cl}} & \equiv & K_{0}+\frac{K_{0}}{4K_{0}-2},\label{eq:K_cl}\\
\delta K_{\text{dis}} & \equiv & \frac{K_{0}^{2}}{2\pi\left(3-2K_{0}\right)}\left(\frac{a_{0}}{l_{e}}\right)\left(\frac{\xi}{2a_{0}}\right)^{\nu},\label{eq:delta_K_dis}
\end{eqnarray}
with $K_{\text{cl}}$ the renormalized Luttinger parameter in the
clean limit, and $\delta K_{\text{dis}}$ the correction due to disorder.
Direct replacement of $K\left(\ell^{*}\right)$ back into Eq. (\ref{eq:S_0_str_coupling})
produces the final Eq. (9) in the manuscript.

\bibliographystyle{apsrev}

\end{document}